\documentclass[aps,pra,showpacs,notitlepage]{revtex4-1}

\usepackage{graphicx}
\usepackage{dcolumn}
\usepackage{bm}

\begin{document}

\title{Experimental evaluation of non-classical correlations between measurement outcomes and target observable in a quantum measurement}

\author{Masataka Iinuma}
 \email{iinuma@hiroshima-u.ac.jp} 
 \homepage{http://home.hiroshima-u.ac.jp/qfg/qfg/index.html}
\author{Yutaro Suzuki}%

\author{Taiki Nii}%

\author{Ryuji Kinoshita}%

\author{Holger F. Hofmann}%

\affiliation{
 Graduate School of Advanced Sciences of Matter, Hiroshima University
 1-3-1 Kagamiyama, Higashi-Hiroshima, 739-8530, Japan
}%

\begin{abstract}
In general, it is difficult to evaluate measurement errors when the initial and final conditions of the measurement make it impossible to identify the correct value of the target observable. Ozawa proposed a solution based on the operator algebra of observables which has recently been used in experiments investigating the error-disturbance trade-off of quantum measurements. Importantly, this solution makes surprisingly detailed statements about the relations between measurement outcomes and the unknown target observable. In the present paper, we investigate this relation by performing a sequence of two measurements on the polarization of a photon, so that the first measurement commutes with the target observable and the second measurement is sensitive to a complementary observable. While the initial measurement can be evaluated using classical statistics, the second measurement introduces the effects of quantum correlations between the non-commuting physical properties. By varying the resolution of the initial measurement, we can change the relative contribution of the non-classical correlations and identify their role in the evaluation of the quantum measurement. It is shown that the most striking deviation from classical expectations is obtained at the transition between weak and strong measurements, where the competition between different statistical effects results in measurement values well outside the range of possible eigenvalues. 
\end{abstract}


\maketitle

\section{Introduction}

Although the uncertainty principle is usually considered to be a fundamental principle of quantum mechanics, its precise theoretical formulation is not always clear. A breakthrough in the investigation of measurement uncertainties was achieved when Ozawa demonstrated in 2003 that the uncertainty trade-off between measurement error and disturbance may be much lower than the uncertainty trade-off between non-commuting properties in a quantum state \cite{Ozawa2003}. Recently, the definitions of measurement uncertainties introduced by Ozawa have been evaluated experimentally using two-level systems such as neutron spins \cite{Erhart2012} and photon polarizations \cite{Baek2013,Rozema2012}. These experimental tests have confirmed the lower uncertainty limits predicted by Ozawa and resulted in the formulation and confirmation of even tighter bounds \cite{Hall2013,Branciard2013,Ringbauer2014,Kaneda2014}. However, there has also been some controversy concerning the role of the initial state in this definition of measurement uncertainties \cite{Watanabe2011,Busch2013,Dressel2014}. It may therefore be useful to take a closer look at the definition of measurement errors and their experimental evaluation.

In principle, it is natural to define the error of a measurement as the statistical average of the squared difference between the measurement outcome and the actual value of the target observable. However, quantum theory makes it difficult to assign a value to an observable when neither the initial state nor the final measurement is represented by an eigenstate of the observable. Nevertheless, the operator formalism defines correlations between the measurement outcome and the operator $\hat{A}$ that represents the target observable, and this correlation between operators can be evaluated by weak measurements \cite{Lund2010} or by statistical reconstruction using variations of the input state \cite{Ozawa2004}. Essentially, the experimental evaluations of Ozawa uncertainties is therefore based on an evaluation of non-classical correlations between the measurement outcome and the target observable in the initial quantum state $\mid \psi \rangle$.

In the following, we investigate the role of non-classical correlations in quantum measurements by applying a sequential measurement to the polarization of a single photon, such that the initial measurement commutes with the target polarization, while the final measurement selects a complementary polarization. In this scenario, the initial measurement can be described by classical error statistics, and the evaluation of the measurement errors corresponds to conventional statistical methods. However, the final measurement introduces non-classical correlations that provide additional information on the target observable. By varying the strength of the initial measurement, we can control the balance between classical and non-classical effects in the correlations. In addition, we obtain two separate measurement outcomes, one of which refers directly to the target observable, and another one which can only relate to the target observable via correlations in the input state. Our measurement results thus provide a detailed characterization of non-classical effects in the relation between measurement outcomes and target observable. In particular, our results show that the intermediate measurement outcome modifies the non-classical correlations between the final outcome and the target observable, which can result in a counter-intuitive assignment of measurement values, where the intermediate measurement outcome and the estimates values seem to be anti-correlated. Our results thus illustrate that the combination of classical and non-classical correlations can be highly non-trivial and should be investigated in detail to achieve a more complete understanding of the experimental analysis of quantum systems.

The rest of the paper is organized as follows. In Sec. \ref{sec:ncorr}, we point out the role of non-classical correlations in the definition of measurement errors and discuss the experimental evaluation using variations of the input state. In Sec. \ref{sec:twolevel}, we derive the evaluation procedure for two level systems and discuss the evaluation of the experimental data. In Sec. \ref{sec:exp}, we introduce the experimental setup and discuss the sequential measurement of two non-commuting polarization components. In Sec. \ref{sec:data}, we discuss the measurement results obtained at different measurement strengths and analyze the role of non-classical correlations in the different measurement regimes. In Sec. \ref{sec:error}, we discuss the effects of non-classical correlations on the statistical error of the measurement. In Sec. \ref{sec:conclusions}, we conclude the paper by summarizing the insights gained from our detailed study of the non-classical aspects of measurement statistics.

\section{Measurement errors and non-classical correlations}
\label{sec:ncorr}

Measurement errors can be quantified by taking the average of the squared difference between the measurement outcomes $A_{\mathrm{out}}(m)$ and the target observable $\hat{A}$. As shown by Ozawa \cite{Ozawa2003}, this definition of errors can be applied directly to the operator statistics of quantum theory, even if the observable $\hat{A}$ does not commute with the measurement outcomes $m$. If the probability of the measurement outcome $m$ is represented by the positive valued operator $\hat{E}_m$, the measurement error for an input state $\mid \psi \rangle$ is given by
\begin{eqnarray}
\label{eq:ozawa}
\varepsilon^2(A) 
& = &  
\sum_{m} \langle \psi \mid (A_m -\hat{A}) \hat{E}_m (A_m -\hat{A}) \mid \psi \rangle
\nonumber \\
& = &
\langle \psi \mid \hat{A}^{2} \mid \psi \rangle + \sum_m A_m^{2} \langle \psi \mid \hat{E}_m \mid \psi \rangle -  2 \sum_{m} A_m \; \Re \left[ \langle \psi   \mid \hat{E}_m \hat{A} \mid \psi \rangle \right].
\end{eqnarray}
The last term in Eq. (\ref{eq:ozawa}) evaluates the correlation between the target observable $\hat{A}$ and the measurement outcome $A_m$. 

If $\hat{A}$ and $\hat{E}_m$ commute, the correlation in Eq. (\ref{eq:ozawa}) can be explained in terms of the joint measurement statistics of the outcomes $m$ and the eigenstate projections $a$, where the eigenvalues of $\hat{E}_m$ determine the conditional probabilities $P(m|a)$ of obtaining the result $m$ for an eigenstate input of $a$. However, the situation is not so simple if $\hat{A}$ and $\hat{E}_m$ do not commute. In this case, an experimental evaluation of the measurement error $\varepsilon(A)^2$ requires the reconstruction of a genuine quantum correlation represented by operator products. Perhaps the most direct method of obtaining the appropriate data is to vary the input state \cite{Ozawa2004}. To obtain the correlation between the measurement outcome $m$ and the observable $\hat{A}$, it is sufficient to use two superposition states as input,
\begin{eqnarray}
\label{eq:pm}
\mid + \rangle = \frac{1}{\sqrt{1+2 \lambda \langle \hat{A} \rangle + \lambda^2 \langle \hat{A}^2 \rangle}} \left(1 + \lambda \hat{A}\right) \mid \psi \rangle
\nonumber \\
\mid - \rangle = \frac{1}{\sqrt{1-2 \lambda \langle \hat{A} \rangle + \lambda^2 \langle \hat{A}^2 \rangle}} \left(1 - \lambda \hat{A}\right) \mid \psi \rangle,
\end{eqnarray}
where the expectation values in the normalization factors refer to the statistics of the original state $\mid \psi \rangle$. It is now possible to determine the correlation between the measurement outcome and the target observable from the weighted difference between the probabilities $P(m|+)$ and $P(m|-)$ obtained with these two superposition states. Specifically,
\begin{equation}
\label{eq:evaluate}
\Re \left[ \langle \psi   \mid \hat{E}_m \hat{A} \mid \psi \rangle \right] = \frac{1}{4 \lambda} \left((1+2 \lambda \langle \hat{A} \rangle + \lambda^2 \langle \hat{A}^2 \rangle) P(m|+) - (1 - 2 \lambda \langle \hat{A} \rangle + \lambda^2 \langle \hat{A}^2 \rangle) P(m|-)\right).
\end{equation}
For $\lambda \ll 1$, the two states correspond to the outputs of a weak measurement with a two level probe state \cite{Hofmann2010}. The variation of input states is therefore closely related to the alternative method of evaluating measurement errors using weak measurements \cite{Lund2010}. 

Eq. (\ref{eq:evaluate}) expresses the correlations between the outcome values $A_m$ and the target observable $\hat{A}$ in terms of conditional expectation values of $\hat{A}$ which correspond to optimal estimates of the target observables,
\begin{equation}
\label{eq:condav}
A_{\mathrm{opt.}}(m) = \frac{\Re \left[ \langle \psi   \mid \hat{E}_m \hat{A} \mid \psi \rangle \right]}{ \langle \psi   \mid \hat{E}_m \mid \psi \rangle}.
\end{equation}
As pointed out by Hall, this optimal estimate is equal to the real part of the weak value conditioned by the post-selection of the measurement outcome $m$ \cite{Hall2004}. If the non-classical correlation in Eq.(\ref{eq:ozawa}) is expressed using the conditional average in Eq.(\ref{eq:condav}), the result reads
\begin{equation}
\varepsilon^2(A) = \langle \hat{A}^2 \rangle - \sum_m \left( A_{\mathrm{opt.}}(m) \right )^2 P(m|\psi) + \sum_m \left( A_m-A_{\mathrm{opt.}}(m))^2 P(m|\psi) \right.
\end{equation}
It is then obvious that the minimal error $\varepsilon^2_{\mathrm{opt.}}(A)$ is obtained for $A_m=A_{\mathrm{opt.}}(m)$, and that this minimal error is given by the difference between the original variance of $\hat{A}$ in the quantum state $\psi$ and the variance of the conditional averages $A_{\mathrm{opt.}}(m)$,
\begin{equation}
\label{eq:varsum}
\varepsilon^2_{\mathrm{opt.}}(A) = \langle \hat{A}^2 \rangle - \sum_m \left(A_{\mathrm{opt.}}(m)\right)^2 P(m|\psi).
\end{equation}
Importantly, all of the necessary information can be obtained experimentally using the superposition input states $\mid+ \rangle$ and $\mid - \rangle$. As will be shown in the following, this means that for two level systems, the non-classical correlations can actually be derived from measurements performed on eigenstates of $\hat{A}$.

\section{Evaluation of two level systems}
\label{sec:twolevel}

In a two level system, all physical properties can be expressed in terms of operators with eigenvalues of $\pm 1$. This results in a significant simplification of the formalism. In particular, it is possible to define the states $\mid + \rangle$ and $\mid - \rangle$ used for the experimental evaluation of non-classical correlations in the measurement errors by setting $\lambda=1$ in Eq. (\ref{eq:pm}). The result is a projection onto eigenstates of $\hat{A}$, so that $\mid + \rangle$ and $\mid - \rangle$ are eigenstates of the target observable $\hat{A}$ with eigenvalues of $+1$ and $-1$, respectively. Surprisingly, this means that the non-classical correlations between measurement outcomes and target observables can be evaluated without applying the measurement of $m$ to the actual input state $\mid \psi \rangle$. According to Eq. (\ref{eq:evaluate}), the relation for the two-level system with eigenvalues of $A_{a}=\pm 1$ and $\lambda=1$ is
\begin{equation}
\label{eq:eval2L}
\Re \left[ \langle \psi \mid \hat{E}_{m} \hat{A} \mid \psi \rangle \right] 
= P(m|+) P(+|\psi) - P(m|-) P(-|\psi).
\end{equation}
Note that this looks like a fully projective measurement sequence, where a measurement of $\hat{A}$ is followed by a measurement of $m$. 
However, such a projective measurement of $\hat{A}$ actually changes the probabilities of the final outcomes $m$. It is therefore quite strange that the correlation between an undetected observable $\hat{A}$ and the measurement result $m$ obtained from an initial state $\psi$ can be derived from a sequential projective measurement, as if the measurement disturbance of a projective measurement of $\hat{A}$ had no effect on the final probabilities of $m$. 

The non-classical features of the correlation in Eq. (\ref{eq:eval2L}) emerge when the conditional average is determined according to Eq. (\ref{eq:condav}),
\begin{equation}
\label{eq:cav2L}
A_{\mathrm{opt.}}(m) = \frac{P(m|+) P(+|\psi) - P(m|-) P(-|\psi)}{P(m|\psi)}.
\end{equation}
Although this equation looks almost like a classical conditional average, it is important to note that the probabilities are actually obtained from two different measurements. As a result, the denominator is not given by the sum of the probabilities in the numerator. In fact, it is quite possible that $P(m|\psi)$ is much lower than the sum of $P(m|+) P(+|\psi)$ and $P(m|-) P(-|\psi)$, so that the conditional average $A_{\mathrm{opt.}}(m)$ is much larger than $+1$ (or much lower than $-1$). In fact, we should expect such anomalous enhancements of the conditional average, since Eq. (\ref{eq:condav}) shows that $A_{\mathrm{opt.}}(m)$ is equal to the weak value of $\hat{A}$ conditioned by $\psi$ and $m$. 

It may seem confusing that the combination of statistical results obtained in two perfectly normal experiments results in the defintion of a seemingly paradoxical conditional average. However, this is precisely why quantum statistics have no classical explanation. In fact, the present two level paradox is simply a reformulation of the violation of Leggett-Garg inequalities \cite{LGI,Knee2012,Suzuki2012}, where it is shown that it is impossible to explain the probabilities $P(m|\psi)$, $P(m|\pm)$ and $P(\pm|\psi)$ as marginal probabilities of the same positive valued joint probability $P(m,\pm|\psi)$. Effectively, the evaluation of measurement errors proposed by Ozawa \cite{Ozawa2004} and applied in the first experimental demonstration \cite{Erhart2012} is identical to the verification of Leggett-Garg inequality violation by parallel measurements proposed in \cite{LGI} and applied in \cite{Knee2012}.

We can now look at the evaluation of the measurement errors in more detail. Using the previous results to express Eq. (\ref{eq:ozawa}) in terms of experimental probabilities, the measurement error is given by
\begin{equation}
\varepsilon^2(A) = 1 + \sum_m A_m^2 P(m|\psi) - 2 \sum_m A_m \left(P(m|+) P(+|\psi) - P(m|-) P(-|\psi) \right).
\end{equation}
Although this is already a great simplification, it is interesting to note that the evaluation used in the first experimental demonstration \cite{Erhart2012} is even more simple. This is because of an additional assumption: if we only allow an assignment of $A_m=\pm1$, so that $m$ can be given by $+$ or $-$ and $A_m^2=1$,  
\begin{equation}
\varepsilon^2(A) = 2 - 2 \left(P(+|+) P(+|\psi) + P(-|-) P(-|\psi) - P(+|-) P(-|\psi) - P(-|+) P(+|\psi)\right).
\end{equation}
In many cases, errors are symmetric, so that $P(+|+)=P(-|-)=1-P_{\mathrm{error}}$ and $P(+|-)=P(-|+)=P_{\mathrm{error}}$. If this assumption is used, the evaluation of measurement errors is completely independent of the input state, since the probabilities of $A_+=+1$ and of $A_-=-1$ add up to one, and the error is simply given by the error observed for eigenstate inputs,
\begin{equation}
\label{eq:error}
\varepsilon^2(A) = 4 P_{\mathrm{error}}.
\end{equation}
Importantly, this result is just a special case where the measurement error appears to be state independent because of a specific choice of $A_m$ for the evaluation of the measurement. In the following, we will consider a setup that explores the optimization of $A_m$ and the role of the non-classical correlations between measurement outcomes and target observable using the evaluation of experimental data developed above.

\section{Sequential measurement of photon polarization}
\label{sec:exp}

As mentioned in the previous section, the anomalous values of the conditional averages $A_{\mathrm{opt.}}(m)$ that also provide the optimal assignments of measurement outcomes $A_m$ originate from the same experimental statistics that are used to violate Leggett-Garg inequalities. We are therefore particularly interested in the correlations between Bloch vector components in the equatorial plane of the Bloch sphere. In the case of photon polarization, these are the linear polarizations, where the horizontal (H) and vertical (V) polarizations define one axis and the diagonal polarizations corresponding to positive (P) and negative (M) superposition of H and V define the orthogonal axis. In terms of operators with eigenvalues of $+1$ and $-1$, these polarizations can be expressed by $\hat{S}_{HV}$ and $\hat{S}_{PM}$. 

If our target observable is $\hat{A}=\hat{S}_{PM}$, any measurement that commutes with $\hat{S}_{PM}$ can be explained in terms of classical statistics. We therefore use a setup that implements a variable strength measurement of diagonal polarization similar to the one we previously used to study Leggett-Garg iequality violations and weak measurements \cite{Suzuki2012,Iinuma2011}. In the output, we then perform a measurement of HV-polarization, so that the total measurement does not commute with the target observable. By dividing the measurement into two parts, we can vary the strength of the non-classical effects and study the transition between classical correlations and quantum correlations in detail.

\begin{figure}[h]
\centerline{\includegraphics[width=70mm]{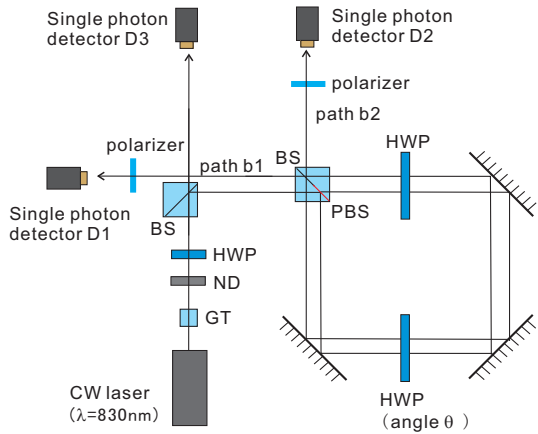}} 
\caption{Experimental setup of the sequential measurement of $\hat{S}_{\mathrm{PM}}$ followed by the projective measurement of $\hat{S}_{HV}$.  This interferometer was realized by using a hybrid cube of a Polarizing Beam Splitter (PBS)  and a Beam Splitter (BS), where the input beam is split by the PBS part and the outputs interfere at the BS part of the cube. The variable strength measurement of the positive (P) and negative (M) superposition of horizontal (H) and vertical (V) polarizations is realized by path interference between the H and the V polarized component. The measurement strength of the PM measurement is controlled by the angle $\theta$ of one of two half-wave plates (HWPs) inside the interferometer, which can be changed from zero for no measurement to 22.5$^{\circ}$ for a fully projective measurement. The other HWP is used for a phase compensation between H and V components.}
\label{fig:setup}
\end{figure}

The experimental setup is shown in Fig. \ref{fig:setup}. As explained in \cite{Iinuma2011}, a variable strength measurement is implemented by separating the horizintal and vertical polarizations at a polarization beam splitter (PBS), rotating the polarizations towards each other using a half-wave plate (HWP) and interfering them at a beam splitter (BS). The effect of the interference is to distinguish P-polarization from M-polarization, where the visibility of the interference and hence the strength of the measurement is controlled by the rotation angle of the HWP, where the angle $\theta$ can be changed from zero for no measurement to 22.5$^{\circ}$ for a fully projective measurement. As shown in Fig. \ref{fig:setup}, the interferomter is a Sagnac type, where the difference between input and output beam splitter is implemented by using a hybrid cube that acts as either a PBS or a BS, depending on the part of the cube on which the beam is incident. Input states were prepared using a Glan-Thompson prism (GT) and another HWP located just before the hybrid cube and a weak coherent light emitted by a CW TiS laser($\lambda=830$ nm).  The output photon numbers in the output paths $b1$ (measurement outcome P or $m_1=+1$) and in the path $b2$ (measurement outcome M or $m_1=-1$) are counted by using the single photon detectors D1 and D2, respectively. Polarizers were inserted to realize the final measurement of $\hat{S}_{HV}$, corrsponding to $m_2=+1$ for H-polarization and $m_2=-1$ for V-polarization. The number of input photons in the initial state was monitored with the single photon detector D3 in order to compensate fluctuations of intensity in the weak coherent light used as input. In the actual setup, we also detected a systematic difference between the reflectivity and the transmissivity of the final BS resulting in a slight change of the orientation of the measurement basis from the directions of PM-polarization. The cancellation of this systematic effect is achieved by exchanging the roles of path b1 and path b2 using the settings of the HWP, which effectively restores the proper alignment of the polarization axes with the measurement \cite{Suzuki2012}.   

The measurement has four outcomes $m=(m_1,m_2)$ given by the combinations of $\hat{S}_{PM}$ eigenvalues ($m_1=\pm1$) and $\hat{S}_{HV}$ eigenvalues ($m_2=\pm 1$). In the absence of experimental errors, the measurement outcomes can be described by pure state projections,
\begin{eqnarray}
\mid +1,+1 \rangle &=& \frac{1}{\sqrt{2}} \left(\cos(2 \theta) \mid H \rangle + \sin(2 \theta) \mid V \rangle  \right),
\nonumber \\
\mid +1,-1 \rangle &=& \frac{1}{\sqrt{2}} \left(\sin(2 \theta) \mid H \rangle + \cos(2 \theta) \mid V \rangle \right),
\nonumber \\
\mid -1,+1 \rangle &=& \frac{1}{\sqrt{2}} \left(\cos(2 \theta) \mid H \rangle - \sin(2 \theta) \mid V \rangle \right),
\nonumber \\
\mid -1,-1 \rangle &=& \frac{1}{\sqrt{2}} \left(\sin(2 \theta) \mid H \rangle - \cos(2 \theta) \mid V \rangle \right).
\end{eqnarray}

\begin{figure}[h]
\begin{center}
    \vspace{0.2cm}
  \includegraphics[width=70mm]{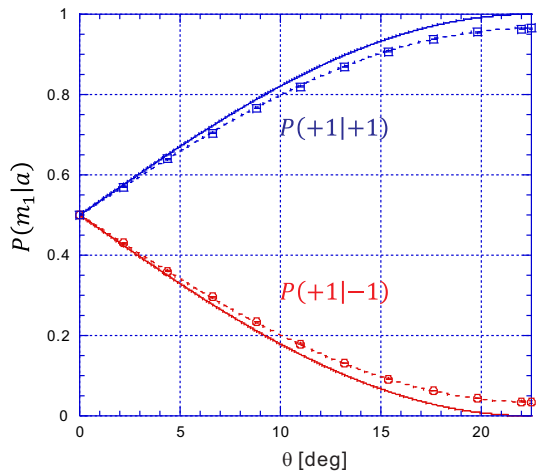}
 \caption{Experimental probabilities $P(m_1|a)$ of the PM-measurement obtained with P polarization ($a=+1$) as the initial state. The solid lines indicate the theoretically expected result for $V_{PM}=1$ and the broken line shows the theoretical expectation for $V_{PM}=0.93$.}  
 \label{fig:PMdata}
\end{center}
\end{figure}

The actual measurement is limited by the visibility of the interferometer, which was independently evaluated as $V_{PM}=0.93$ at $\theta=22.5^{\circ}$. It is possible to characterize the measurement error of the PM-measurement by preparing P-polarized and M-polarized input photons. If $A_m=+1$ is assigned to the $m_1=+1$ outcomes, and $A_m=-1$ is assigned to the $m_1=-1$ outcomes, this corresponds to a measurement of the error probability $P_{\mathrm{error}}$ in Eq.(\ref{eq:error}),
\begin{equation}
P_{\mathrm{error}} = P(m_1=-a|a) = \frac{1}{2}\left(1 - V_{PM} \sin(4 \theta) \right).
\end{equation}
Fig. \ref{fig:PMdata} shows the experimental results obtained with our setup. Note that this graph also provides all of the data needed to determine the probabilities $P(m_1,m_2|a)$ for the analysis of the conditional averages $A_{\mathrm{opt.}}(m)$ in the following section, since $P(m_1,m_2|a)=P(m_1|a)/2$. 

For completeness, we have also evaluated the experimental errors in the final measurement of HV-polarization. We obtain a visibility of $V_{HV}=0.9976$ for the corresponding eigenstate inputs. With this set of data, we can fully characterize the performance of the measurement setup, as shown in the analysis of the following experimental results.

\section{Experimental evaluation of non-classical correlations}
\label{sec:data}

\begin{figure}[h]
\begin{center}
    \vspace{0.2cm}
  \includegraphics[width=70mm]{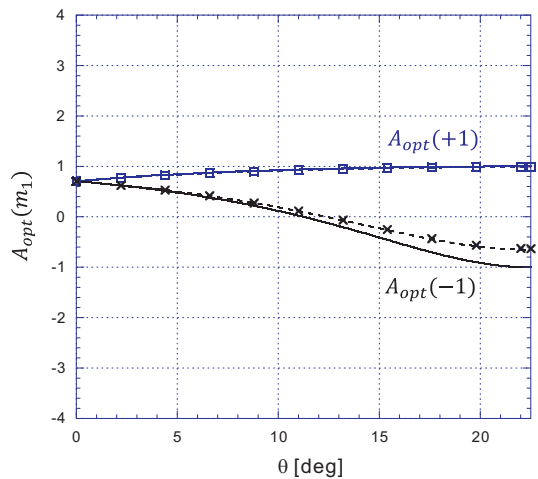}
 \caption{Conditional average $A_{\mathrm{opt.}}(m_1)$ of the PM-polarization $\hat{S}_{PM}$ obtained after a measurement of $m_1=+1$ (P-polarization) or $m_1=-1$ (M-polarization) at different measurement strengths $\theta$. At $\theta=0$, the measurement outcome is random ($P_{\mathrm{error}}=1/2$ and the conditional average is simply given by the original expectation value of the input state. As the likelihood of measurement errors decreases, the conditional average approaches the value given by the measurement outcome $m_1$.}
 \label{fig:Bayes}
\end{center}
\end{figure}

\begin{figure}[h]
\begin{center}
    \vspace{0.2cm}
  \includegraphics[width=70mm]{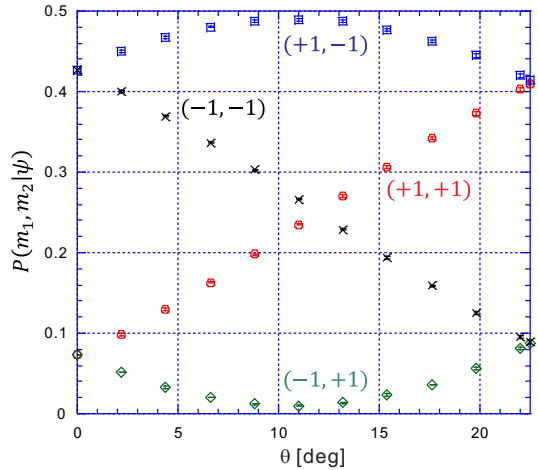}
 \caption{Probabilities $P(m_1,m_2|\psi)$ for the outcomes of the sequential measurement of $m_1$ (PM-polarization) and $m_2$ (HV-polarization) on an input state polarized at 67.5$^{\circ}$, halfway between $P$ and $V$.}
 \label{fig:ALLdata}
\end{center}
\end{figure}

To obtain non-classical correlations between $\hat{S}_{PM}$ and $\hat{S}_{HV}$, we chose an input state $\psi$ with a linear polarization at 67.5$^{\circ}$, halfway between the P-polarization and the V-polarization. For this state, the initial expection value of the target observable is
\begin{equation}
\langle \hat{S}_{PM} \rangle = \frac{1}{\sqrt{2}}.
\end{equation}
We can now start the analysis of measurement errors by considering only the outcome $m_1$, in which case the measurement operators $\hat{E}_m$ commute with the target observable and the problem could also be analyzed using classical statistics. Specifically, commutativity means that the probability $P(m_1|\psi)$ is unchanged if a projective measurement of $\hat{S}_{PM}$ is performed before the measurement of $m_1$. It is therefore possible to determine $P(m_1|\psi)$ from the conditional probabilities $P(m_1|a)$ and $P(a|\psi)$, which results in a classical conditional average for $\hat{A}=\hat{S}_{PM}$ given by
\begin{eqnarray}
\label{eq:Bayes}
A_{\mathrm{opt.}}(m_1) &=& \frac{P(m_1|+) P(+|\psi) - P(m_1|-) P(-|\psi)}{P(m_1|+) P(+|\psi) + P(m_1|-) P(-|\psi)}.
\nonumber \\
&=& \frac{(1-2 P_{\mathrm{error}}) m_1 + \langle \hat{S}_{PM} \rangle}{m_1 +(1-2 P_{\mathrm{error}}) \langle \hat{S}_{PM} \rangle} \; m_1.
\end{eqnarray}
Eq. (\ref{eq:Bayes}) shows that the conditional averages are found somewhere between the original expectation value of $\langle \hat{S}_{PM} \rangle$ for $P_{\mathrm{error}}=1/2$ and the measurement result $m_1$ for $P_{\mathrm{error}}=0$. In the experiment, the error probability is controlled by the measurement strength $\theta$ as shown in Fig. \ref{fig:PMdata}. The corresponding dependence of $A_{\mathrm{opt.}}(m_1)$ on $\theta$ is shown in Fig. \ref{fig:Bayes}. 
It should be noted that the result does not change if it is based on the joint probabilites $P(m_1,m_2|\psi)$ shown in Fig. \ref{fig:ALLdata}, since the marginal probabilities $P(m_1|\psi)$ of this joint probability distribution are equal to the sums of the sequential measurement probabilities $P(m_1|a) P(a|\psi)$. This is an important fact, since the actual value of $a$ is fundamentally inaccessible once the final measurement of $m_2$ is performed,  regardless whether the data obtained from $m_2$ is used or not. Even though the correlation between $\hat{S}_{PM}$ and $m_1$ can be explained using classical statistics, this possibility does not imply that we can safely assign a physical reality $a$ to the observable. The distinction between classical and non-classical correlations is therefore more subtle than the choice of measurement strategy. 

Up to now, the analysis does not include any non-classical correlations, since the measurement is only sensitive to the target observable $\langle \hat{S}_{PM} \rangle$. This situation changes if we include the outcome $m_2$ of the final HV-measurement in the evaluation of the experimental data. Importantly, we intend to use the information gained from the outcome of the HV-measurement to update and improve our estimate of the PM-polarization in the input. For that purpose, we need to evaluate the non-classical correlations between $\langle \hat{S}_{PM} \rangle$ and $\langle \hat{S}_{HV} \rangle$, which can be done using the method developed in section \ref{sec:twolevel}. In addition to the known probabilities $P(a|\psi)$ and $P(m_1,m_2|a)$, we now need to include the measurement outcomes $P(m_1,m_2|\psi)$ which provide the essential information on the non-classical correlations. The experimental results for $P(m_1,m_2|\psi)$ obtained at variable measurement strengths $\theta$ are shown in Fig. \ref{fig:ALLdata}. The question is how the final result $m_2$ changes our estimate of $\hat{S}_{PM}$. According to Eq. (\ref{eq:cav2L}), we can find the answer by dividing the difference between the probabilities of a measurement sequence of $a$ followed by $(m_1,m_2)$ by the probabilities obtained by directly measuring $(m_1,m_2)$, 
\begin{eqnarray}
\label{eq:seq}
A_{\mathrm{opt.}}(m_1,m_2) &=& \frac{P(m_1,m_2|+) P(+|\psi) - P(m_1,m_2|-) P(-|\psi)}{P(m_1,m_2|\psi)}
\nonumber \\ &=&
\frac{m_1 (1-2 P_{\mathrm{error}}) + \langle \hat{S}_{PM} \rangle}{4 P(m_1,m_2|\psi)}.
\end{eqnarray}
Note that the simplification of this relation is possible because the result $m_2$ of the HV-measurement is completely random when the input states are eigenstates of PM-polarization, so that $P(m_1,m_2|\pm) = P(m_1|\pm)/2$. Thus the $m_2$-dependence of the conditonal average only appears in the denominator. Specifically, the difference in the probability of finding H-polarization ($m_2=+1$) or V-polarization ($m_2=-1$) in the final measurement translates directly into a difference in the conditional probabilities, where a lower probability of $m_2$ enhances the estimated value $A_{\mathrm{opt.}}(m_1,m_2)$. 

\begin{figure}[h]
\begin{center}
    \vspace{0.2cm}
  \includegraphics[width=70mm]{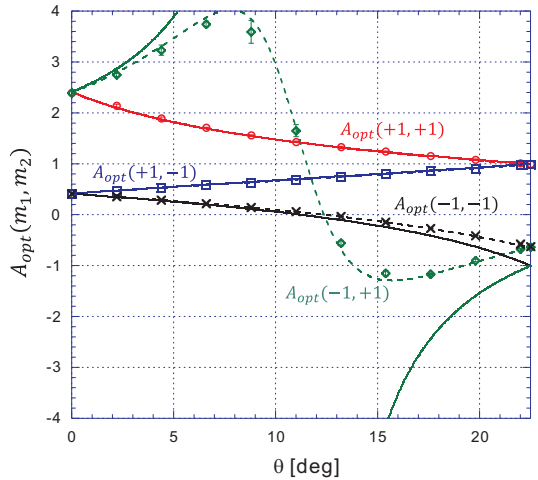}
 \caption{Conditional averages $A_{\mathrm{opt.}}(m_1,m_2)$ as a function of measurement strength $\theta$. The solid curve represents the theoretical prediction for a measurement without experimental imperfections, the broken line was calculated for an interferomter visibility of $V_{PM}=0.93$.}
 \label{fig:Aopt}
\end{center}
\end{figure}

Fig. \ref{fig:Aopt} shows the dependence of the conditional averages of $\hat{S}_{PM}$ on the measurement strength $\theta$. Significantly, the low probabilities of finding H-polarization ($m_2=+1$) result in estimates of $A_{\mathrm{opt.}}(m_1,m_2)$ that lie outside of the range of eigenvalues. The difference between $A_{\mathrm{opt.}}(+1,+1)$ and $A_{\mathrm{opt.}}(+1,-1)$ corresponds to the contribution of the non-classical correlation between $\hat{S}_{PM}$ and $m_2$, whereas the difference between $A_{\mathrm{opt.}}(+1,-1)$ and $A_{\mathrm{opt.}}(-1,-1)$ corresponds to the contribution of the correlation between $\hat{S}_{PM}$ and $m_1$, which is closely related to the classical correlation that determines the behavior of $A_{\mathrm{opt.}}(m_1)$ in Fig. \ref{fig:Bayes}. As the measurement strength increases, the correlation between $\hat{S}_{PM}$ and $m_2$ drops towards zero and the correlation between $\hat{S}_{PM}$ and $m_1$ increases, approaching the ideal identification of the measurement outcome $m_1$ with the eigenvalue of $\hat{S}_{PM}$. For intermediate measurement strengths, it is important to consider the correlations between the measurement outcomes as well, indicating that the non-classical correlations associated with $m_2$ are modified by the results of $m_1$ and vice versa. The adjustment of measurement strength is therefore a powerful tool for the analysis of masurement statistics that may give us important new insights into the way that classical and non-classical correlations complement each other. 

The conditional average $A_{\mathrm{opt.}}(m_1,m_2)$ is obtained from the correlations between $\hat{S}_{PM}$ and the two measurement results $m_1$ and $m_2$ that originate from the statistics of the initial state $\psi$. Specifically, the estimate is obtained by updating the initial statistics of $\psi$ based on the outcomes $m_1$ and $m_2$, where the measurement strength controls the relative statistical weights of the information obtained from $m_1$ and $m_2$. At a maximal measurement strength of $\theta=22.5^\circ$, the PM-measurement completely randomizes the HV-polarization, so that the conditional average $A_{\mathrm{opt.}}(m_1,m_2)$ is independent of $m_2$ and the estimation procedure is based on the classical correlations between $m_1$ and $\hat{S}_{PM}$. As the measurement strength is weakend, a small contribution of non-classical correlations emerges as the conditional averages for $m_2=+1$ and for $m_2=-1$ split, with the estimates for the more likely $m_2$-outcomes dropping towards zero and the estimates for the less likely $m_2$-outcomes diverging to values greater than $+1$ for $m_1=+1$ and more negative than $-1$ for $m_1=-1$. Even small contributions of non-classical correlations therefore result in estimates that cannot be reproduced by classical statistics. Due to experimental imperfections, the anomalous values of $A_{\mathrm{opt.}}(+1,+1)>1$ are easier to observe than the anomalous values of $A_{\mathrm{opt.}}(-1,+1)<-1$. Specifically, the small probabilities of the result $(-1,+1)$ are significantly enlarged by the noise background associated with limited visibilities. As the measurement strength drops, the initial bias in favor of P-polarization in the input state $\psi$ begins to outweigh the effect of the measurement result of $m_1=-1$ that would indicate M-polarization. Of particular interest is the crossing point around $\theta=12.3^\circ$, where the initial information provided by $\psi$ and the measurement information $m_1$ become equivalent and the estimate is $A_{\mathrm{opt.}}(-1,m_2)=0$ for both $m_2=+1$ and $m_2=-1$. For measurement strengths below this crossing point, the initial bias provided by the initial state towards P-polarization clearly dominates the estimate, resulting in positive values of $A_{\mathrm{opt.}}(-1,m_2)$. Significantly, the increase of the estimate with reduction in measurement strength is much faster for $m_2=+1$ than for $m_2=-1$, since the lower probability of the outcome $m_2=+1$ effectively enhances the statistical weight of the information. For $\theta \approx 11^\circ$, this enhancement of the estimate even results in a crossing between $A_{\mathrm{opt.}}(-1,+1)$ and $A_{\mathrm{opt.}}(+1,+1)$, so that the value estimated for an outcome of $m_1=-1$ actually exceeds the value estimated for an outcome of $m_1=+1$ at measurement strengths of $\theta < 11^\circ$. This counter-intuitive difference between the outcome of the PM-measurement and the estimated value of PM-polarization appears due to the effects of the measurement outcome $m_1$ on the quantum correlations between $m_2$ and the target observable $\hat{S}_{HV}$ in the initial state. Specifically, low probability outcomes always enhance the correlations between measurement results and target observable. Therefore, the low probability outcome $m_1=-1$ enhances the correlation between $m_2=+1$ and $\hat{S}_{HV}$, which favours the P-polarization. On the other hand, the much higher probability of $m_1=+1$ does not result in a comparative enhancement of this correlation, so that the estimated value $A_{\mathrm{opt.}}(+1,+1)$ for an outcome of $m_1=+1$ is actually lower than the estimated value $A_{\mathrm{opt.}}(-1,+1)$ for an outcome of $m_1=-1$. These non-classical aspects of correlations between measurement results and target observable highlight the importance of the relation between the two measurement outcomes: it is impossible to isolate the measurement result $m_1$ from the context established by both $\psi$ and $m_2$. Since the estimated values $A_{\mathrm{opt.}}(m_1,m_2)$ correspond to weak values, this observation may also provide a practical example of the relation between weak values and contextuality \cite{Pusey2014}. 

In the limit of zero measurement strength ($\theta=0$), the estimated values depend only on $m_2$, with the unlikely measurement outcome of $m_2=+1$ resulting in an anomalous weak value of $A_{\mathrm{opt.}}(m_1,+1)=\sqrt{2}+1$ and the likely outcome of $m_2=-1$ resulting in a weak value estimate of $A_{\mathrm{opt.}}(m_1,-1)=\sqrt{2}-1$. Since these estimates are based only on the outcomes of precise measurements of HV-polarization, they provide a direct illustration of the non-classical correlation between $\hat{S}_{PM}$ and $\hat{S}_{HV}$ in $\psi$. Due to the specific choice of initial state, $A_{\mathrm{opt.}}(m_1,+1)$ is larger than $A_{\mathrm{opt.}}(m_1,-1)$, which means that the detection of H-polarization makes P-polarization more likely, while the detection of V-polarization increases the likelihood of M-polarization. If we disregard for a moment that the estimated values for $m_2=+1$ lie outside the range of possible eigenvalues, we can give a fairly intuitive characterization of this non-classical correlation. Clearly, the lowest likelihood is assigned to the combination of H-polarization and M-polarization, which are the least likely polarization results obtained in separate measurements of HV-polarization and PM-polarization for the input state $\psi$. We can therefore summarize the result by observing that quantum correlations between Bloch vector components 
strongly suppress the joint contributions of the least likely results, to the point where the correlation can exceed positive probability boundaries, corresponding to an implicit assignment of negative values to the combination of the two least likely outcomes \cite{Suzuki2012}.

The results presented in this section clearly show that the final HV-measurement provides additional information about the target observable $\hat{A}=\hat{S}_{PM}$. We can therefore expect that the measurement error will be reduced significantly if we use $A_{m_1,m_2}=A_{\mathrm{opt.}}(m_1,m_2)$ as measurement result assigned to the joint outcome $(m_1,m_2)$. In the final section of our discussion, we will therefore take a look at the measurement errors obtained at different measurement strengths $\theta$ and identify the amount of PM-information obtained from the measurement of HV-polarization. 

\section{Evaluation of measurement errors}
\label{sec:error}

\begin{figure}[h]
\begin{center}
    \vspace{0.2cm}
  \includegraphics[width=70mm]{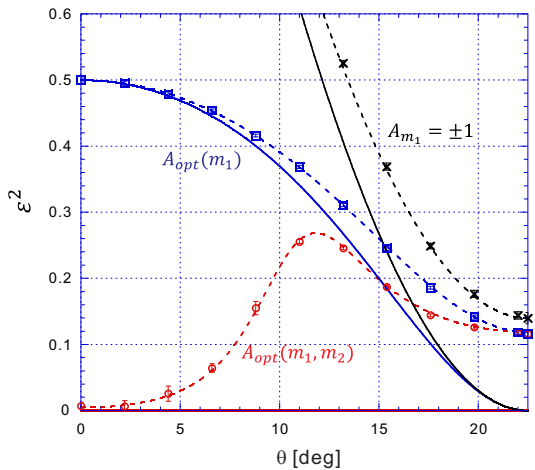}
 \caption{Measurement errors for different measurement strategies. The highest errors are obtained by assigning eigenvalues of $A_{m_1}=\pm 1$ to the outcomes $m_1$ of the PM-measurement. Optimization of the estimate based on $m_1$ results in an error that decreases with increasing measurement strength. By basing the estimate on the combined outcomes $(m_1,m_2)$, it is possible to achieve errors close to zero for low measurement strength $\theta$, since the undisturbed HV-measurement provides maximal information on the PM-polarization through the non-classical correlations between $\hat{S}_{PM}$ and $\hat{S}_{HV}$ in the initial state $\psi$.}
 \label{fig:results}
\end{center}
\end{figure}

According to Eq. (\ref{eq:varsum}), the measurement errors for optimized measurement outcomes $A_{m}=A_{\mathrm{opt.}}(m)$ can be evaluated directly by subtracting the statistical fluctuations of $A_{m}$ from the initial fluctuations of the target observable $\hat{A}$ in the initial state $\psi$. We can therefore use the results of the previous sections to obtain the measurement errors $\varepsilon^2(A)$ for the measurement outcomes $m_1$ and for the combined measurement outcomes $(m_1,m_2)$. The results are shown in Fig. \ref{fig:results}, together with the measurement error given by Eq. (\ref{eq:error}), which is obtained by assigning values of $A_{m_1}=\pm 1$ to the measurement outcomes $m_1$.

Not surprisingly, the sub-optimal assignment of eigenvalues to the measurement outcomes results in much avoidable extra noise. In fact, the error for this assignment exceeds the uncertainty of $\Delta A^2=0.5$ for the initial state $\psi$ at measurement strengths of $\theta < 13.5^\circ$, indicating that one can obtain a better estimate of PM-polarization from the expectation value of the input state. 
This never happens in the case of the errors $\varepsilon_{\mathrm{opt.}}$ associated with the optimal estimates of the target observable, since the optimized estimates based on the conditional averages for the different measurement outcomes include the information of the initial state. In the case of the classical estimate $A_{\mathrm{opt.}}(m_1)$ obtained from the variable strength PM-measurement, the measurement error drops gradually from the variance of the initial state at $\theta=0$ to a residual error caused by the limited visibility $V_{PM}$ at $\theta=22.5^\circ$. By including the information of the final HV-measurement, the estimate can be improved to $A_{\mathrm{opt.}}(m_1,m_2)$, resulting in a reduction of the error that is particularly significant when the measurement strength approaches $\theta=0$. 

The most interesting experimental result is definitely the error obtained for the optimal estimate $A_{\mathrm{opt.}}(m_1,m_2)$, which summarizes all of the available information in the estimates shown in Fig. \ref{fig:Aopt}. Theoretically, the error of this estimate would be zero if the measurements could be performed without any experimental imperfections, as indicated by the red solid line in Fig. \ref{fig:results}. The actual results are close to zero error in the limit of low measurement strength. In this limit, the high visibility of the final HV-measurement for $m_2$ dominate the estimate, with a much lower impact of the less reliable PM-measurement
for $m_1$. The errors then start to rise as the experimental values of $A_{\mathrm{opt.}}(-1,+1)$ in Fig. \ref{fig:Aopt} reach their maximal values near $\theta=8^\circ$. 
The value of the error continues to rise beyond the maximum of $A_{\mathrm{opt.}}(-1,+1)$ and reaches its maximal value near the $\theta=12.3^\circ$ crossing point where $A_{\mathrm{opt.}}(-1,+1)=A_{\mathrm{opt.}}(-1,-1)=0$. At this point, the estimate is particularly sensitive to measurement noise, since the extremely low probabilities of an outcome of  $(-1,+1)$ are strongly affected by experimental noise backgrounds.  For measurement stengths greater than this crossing point ($\theta>12.3^\circ$), the error of $A_{\mathrm{opt.}}(m_1,m_2)$ is not much lower than the error of $A_{\mathrm{opt.}}(m_1)$, indicating that the final measurement result $m_2$ provides only very little additional measurement information on $\hat{S}_{PM}$. This appears to be a result of the experimental noise in the PM-measurement, which limits the error to $\varepsilon^2=0.12$ at a maximal measurement strength of $\theta=22.5^\circ$. 

In summary, the analysis of the measurement errors shows that the non-classical correlation between $m_2$ and $\hat{S}_{PM}$ used to obtain the estimate $A_{\mathrm{opt.}}(m_1,m_2)$ in the limit of weak measurement interactions results in much lower errors than the use of the classical correlations between $m_1$ and $\hat{S}_{PM}$ that dominate in the strong measurement regime. This is a result of the fact that the errors in the limit of weak measurement are dominated by the HV-visibility of the setup, while the errors in the strong measurement regime mostly originate from the PM visibility, which happens to be much lower than the HV-visibility in the present setup.
Our setup is therefore ideally suited to illustrate the importance of non-classical correlations in the evaluation of measurement errors when the initial state is taken into account. The optimal estimate $A_{\mathrm{opt.}}(m_1,m_2)$ is obtained by 
considering the specific relation between the measurement outcomes and the target observable in the specific input state, which may result in counter-intuitive assignments of values to the different measurement outcomes. In the present case, the lowest errors are obtained as a consequence of this counter-intuitive assignment, since the experimental setup is particularly robust against experimental imperfections 
in the regime of low measurement strength which is most sensitive to the effects of non-classical correlations. Our results thus provide a particularly clear experimental demonstration of the reduction of measurement errors by non-classical correlations between measurement result and target observable in the initial quantum state.

\section{Conclusions}
\label{sec:conclusions}

We have investigated the non-classical correlations between the outcomes of a quantum measurement and the target observable of the measurement by studying the statistics of measurement errors in a sequential measurement. In the initial measurement, the measurement operator commutes with the target observable and the measurement outcome $m_1$ relates directly to the target observable, while the final measurement of a complementary observable introduces the effect of non-classical correlations between the outcome $m_2$ and the target observable. To evaluate the errors, we applied the operator formalism introduced by Ozawa and show that the evaluation of two-level statistics can be performed by combining the measurement statistics of the input state $\psi$ with the statistics obtained from eigenstate inputs of the target observable. By combining the statistics of separate measurements according to the rules obtained from the operator formalism, it is possible to identify the optimal estimate of the target observable using only the available experimental data. Due to the specific combination of the statistical results, this estimate can exceed the limits of classical statistics by obtaining values that lie outside the range of possible eigenvalues. Typically, the least likely outcomes are associated with extreme values of the target observable. In the present experiment, we find extremely high estimates of the target observable when the strength of the initial measurement is weak and the measurement result is dominated by the non-classical correlations between the target observable and the complementary observable detected in the final measurement. In this limit, the initial measurement outcome that refers directly to the target observable mainly enhances or reduces the effects of the non-classical correlations, which results in the counter-intuitive anti-correlation between the actual measurement result and the associated estimate of the target observable for a final outcome of $m_2=+1$. 

Our discussion provides a more detailed insight into the experimental analysis of measurent errors that has recently been used to evaluate the uncertainty limits of quantum measurements derived by Ozawa \cite{Ozawa2003,Erhart2012,Baek2013,Rozema2012,Hall2013,Ringbauer2014,Kaneda2014}. It is important to note that the estimation procedure associated with this kind of error analysis also reveals important details of the non-classical statistics originating from the correlations between physical properties in the initial state. In the present work, we have taken a closer look at the experimental analysis of measurement errors and clarified its non-classical features. The results show that some of the effects involved in the optimal evaluation of the experimental data are rather counter-intuitive and exhibit features that exceed the possibilities of classical statistics in significant ways. For a complete understanding of measurement statistics in quantum mechanics, it is therefore necessary to explore the effects of non-classical correlations in more detail, and the present study may be a helpful starting point for a deeper understanding of the role such correlations can play in various measurement contexts.
 
\section*{Acknowledgments}
This work was supported by JSPS KAKENHI Grant Number 24540428. One of authors (Y.S.) is supported by Grant-in-Aid for JSPS Fellows 265259.

\end{document}